# Strong superradiance of coherently coupled magnetic dipole emitters mediated by whispering gallery modes of a subwavelength all-dielectric cavity

Ma-Long Hu, Xiao-Jing Du, Lin Ma, Jun He and Zhong-Jian Yang*

*Hunan Key Laboratory of Nanophotonics and Devices, School of Physics and Electronics, Central South University, Changsha 410083, China*

*E-mail: zjyang@csu.edu.cn

**Abstract**: The interaction of magnetic dipole (MD) emitters and common photonic cavities is usually weak, which is partially due to the low magnetic near field enhancements of the cavities. Here, we show that whispering gallery modes (WGMs) of a subwavelength dielectric cavity can not only greatly boost the emission rate of a MD emitter but also bring efficient couplings between coherent MD emitters. In a WGM cavity, the maximal emission rate ($\gamma_{\max}$) of a single emitter occurs at an antinode of the field pattern. The emission of the MD emitter can also be greatly affected by another coherent one depending on the magnetic field response of the WGM. The maximal contribution can also reach $\gamma_{\max}$. Notably, the cooperative emission rate of the coherent MD emitters does not decay with distance in the considered range due to the high-quality feature of a WGM. In contrast to the emission, the absorption of an emitter is hardly affected by the coherent couplings between emitters mediated by a WGM. The difference between the performances of emission and absorption is highly related to the excitation behaviors of WGMs. Our results are important for enhanced magnetic light-matter interactions.

# I. INTRODUCTION

Modifying the spontaneous emission rate of quantum emitters is a fundamental aspect of the light-matter interactions area, which can find many state-of-the-art applications such as single-photon sources[1,2], fluorescent microscopy and nanoscale imaging[3-5], subwavelength semiconductor lasers[6,7]. Up to the present day, there have been a large number of excellent works devoted to the discussion of the spontaneous emission enhancement from electric dipole (ED) transitions [8-14], while only a few works have paid close attention to the magnetic dipole (MD) counterpart [15-17]. This is due to the fact that commonly the interaction of matter with the magnetic field of light is approximately five orders of magnitude weaker as compared to the interaction of matter with the electric field of light[18,19]. However, some rare-earth ions, such as $Eu^{3+}$ and $Er^{3+}$, exhibit strong MD transitions which can even be comparable to the ED transitions [20-24]. Thus, they have drawn increasing attention in the aspect of the interactions between matter and the magnetic field of light [25,26]. Those interactions can find significant applications, for example, white light-emitting diodes (LEDs) [27,28].

The interaction between quantum emitters and local electromagnetic fields can be enhanced by tailoring the local density of optical states (LDOS), e.g., by means of various types of optical cavities including plasmonic nanocavities [8,15,29-31], all-dielectric photonic cavities [32-37]. Plasmonic nanocavities can provide high electric near field confinement, and have been widely used to turn the interactions of ED emitters and light. Plasmonic nanocavities can also be used to tailor the magnetic light-matter interactions [15,38,39]. However, they have several shortcomings referring to their coupling with MD emitters. The first one is the efficient magnetic field can only be found in complex geometries which need careful preparation [15]. Another shortcoming is that they have material losses. Recent years, all-dielectric subwavelength cavities with high refractive index (e.g. silicon) have drawn lots of interests as they exhibit strong far-field electromagnetic resonances while their material losses are low or ignorable [40,41]. Those resonant modes are accompanied by corresponding electric and magnetic near field enhancements. These properties also make them possible platforms for both electric[42,43] and magnetic light-matter interactions [17,41,44,45]. The electromagnetic near filed enhancements of reported common modes provided by all-dielectric cavities are not high enough (~$10^1$). Those modes includes a magnetic dipole mode [40,41], anapole and quasi-bound states in the continuum (quasi-BIC) modes [46-49]. Recently, it was been shown that whispering gallery modes (WGMs) of subwavelength dielectric cavities can support ultrahigh field enhancement [35] due to their well-known high-quality ($Q$) factors. Their electric and magnetic field enhancements are more than one order of magnitude larger than that of the above modes, respectively. Thus, they hold the possibility to enable strong magnetic light-matter interactions [33]. Note that WGMs have been widely shown in dielectric microcavities [50,51] whose size is much larger than the working length. Using WGMs to enhance the electric light-matter interactions has been widely studied [52,53].

The emission property of a quantum emitter can also be modified by its interaction with other quantum emitters, which is usually termed the Dicke effect [10,54,55]. In vacuum, the decay rate of a single emitter is $\gamma_0$, the contribution of the decay rate from another one can reach $\gamma_0$ when they are close to each other and the contribution decays fast with their distance [56]. By introducing a photonic cavity, it is well known that the decay rate of an emitter $\gamma$ can be greatly enhanced (large $\gamma/\gamma_0$) [9,10]. The contribution from another emitter to the above emitter can also reach ~ $\gamma/\gamma_0$ due to the enhanced cooperative interaction mediated by the photonic cavity. Besides, the cooperative effect strength can decay slower compared to that in vacuum [10,52]. A typical system is a plasmonic waveguide hybrid system where the $\gamma/\gamma_0$ can reach ~$10^4$[10]. On the other hand, the plasmonic waveguide still suffers from high material loss [10,54,57,58]. Furthermore, the previous investigations are mainly based on the ED emitters. With the development of the enhanced magnetic light-matter interactions [17,33,45,59], the cooperative effect between MD emitters mediated by proper magnetically resonant cavities becomes more realistic. However, this has not drawn enough attention.

Here, we theoretically study the radiation properties of coherently interacting identical MD emitters mediated by a subwavelength dielectric disk cavity of high-$Q$ WGMs. The material of the disk is taken as silicon with the refractive index $n = 3.5$. The efficient coupling between MD emitters and a subwavelength WGM can induce a high emission rate enhancement ($\sim 10^4$) of a single emitter and efficient couplings between MD emitters. The contribution to the emission of an emitter from another one is highly dependent on the relative phase difference between them and the magnetic field enhancement provided by the WGM. When one MD emitter is located at an antinode of the field pattern of a WGM, which corresponds to the maximal emission rate ($\gamma_{\max}$) of a single emitter, the maximal contribution from another coherent MD emitter can also reach $\gamma_{\max}$. Based on this fact, the emission of an MD emitter can be largely enhanced with many other coherent emitters. Due to the high-$Q$ feature of a WGM, the cooperative emission rate of the coherent MD emitters does not decay with distance in our considered range. This is a significant advantage over the plasmonic system where the cooperative emission suffers from high losses. We also consider the effect of the coherence between MD emitters on the absorption (or extinction) properties of the emitters in such a hybrid system. However, the absorption of an emitter is weakly affected by the coherent couplings between emitters mediated by a WGM. The difference between the performances of the emission and absorption behaviors is highly related to the fact that the WGMs involved are excited differently.

## II. THEORETICAL APPROACH

We consider a coherent hybrid system consisting of a dielectric resonant cavity and two identical two-level MD emitters [Fig. 1(a)]. The Hamiltonian for the emitters in the coherent system can be expressed as [33,56]

$$H_m = 2\hbar\omega_0 \hat{a}^\dagger \hat{a} - \sum_{i=1}^{2}\left(\mu_m B_i \hat{a} + \mu_m B_i^* \hat{a}^\dagger\right), \tag{1}$$

where $\mu_m$ is the MD transition matrix element of each emitter, $B_1$ and $B_2$ are the magnetic fields felt by the first (MD1) and second MD emitter (MD2), respectively.

In the hybrid system, $B_1$ can be written as $B_1 = B_0 + X_1 m_{cav}$, where $B_0$ is the magnetic field of the excitation plane wave, $X_1 m_{cav}$ represents the contribution from the resonant cavity [33,59]. $m_{cav}$ is the high-order magnetic moment of the cavity, and $X_1$ represents the proportional coefficient between the magnetic field produced by the resonant cavity and $m_{cav}$. Similarly, $B_2$ can be written as $B_2 = B_0 + X_2 m_{cav}$, $X_2$ is the proportional coefficient at the location of the MD2. The high-order magnetic moment of the cavity can be expressed as $m_{cav} = \mu_0^{-1} B_{cav} \alpha_M$ by analogy with the magnetic dipole moment, where $B_{cav}$ is the total magnetic field felt by the resonant cavity, $\alpha_M$ is the high-order magnetic polarizability of the cavity, and $\mu_0$ is the permeability of vacuum. $B_{cav}$ is contributed from the excitation plane wave $B_0$ and the two emitters $B_{cav} = B_0 + Y_1 m_1 + Y_2 m_2$, where $m_1 = \mu_m \left( \rho_{12}^{(1)} + \rho_{21}^{(1)} \right)$ and $m_2 = \mu_m \left( \rho_{12}^{(2)} + \rho_{21}^{(2)} \right)$ are the magnetic dipole moments of MD1 and MD2, respectively; $Y_1$ and $Y_2$ are the proportional coefficient of MD1 and MD2, respectively. Based on the density matrix, the magnetic dipole moments of the MD emitters can also be written as

$$m_1 = \mu_m \left( \bar{\rho}_{12}^{(1)} e^{i\omega t} + \bar{\rho}_{21}^{(1)} e^{-i\omega t} \right), \ m_2 = \mu_m \left( \bar{\rho}_{12}^{(2)} e^{i\omega t} + \bar{\rho}_{21}^{(2)} e^{-i\omega t} \right). \tag{2}$$

The magnetic field of incident light as a function of time can be written as $B(t) = (B_0/2) e^{i\omega t} + (B_0/2) e^{-i\omega t}$, Thus, the magnetic field $B_{cav}$ can be expressed as

$$B_{cav}(t) = \left( \frac{B_0}{2} + Y_1 \mu_m \bar{\rho}_{21}^{(1)} + Y_2 \mu_m \bar{\rho}_{21}^{(2)} \right) e^{-i\omega t} + \left( \frac{B_0}{2} + Y_1 \mu_m \bar{\rho}_{12}^{(1)} + Y_2 \mu_m \bar{\rho}_{12}^{(2)} \right) e^{i\omega t}. \tag{3}$$

Therefore, the magnetic field felt by MD1 can be written as

$$B_1(t) = \frac{B_0}{2} e^{-i\omega t} + \frac{B_0}{2} e^{i\omega t} + X_1 \mu_0^{-1} \alpha_M \left[ \left( \frac{B_0}{2} + Y_1 \mu_m \bar{\rho}_{21}^{(1)} + Y_2 \mu_m \bar{\rho}_{21}^{(2)} \right) e^{-i\omega t} \right.$$
$$\left. + \left( \frac{B_0}{2} + Y_1 \mu_m \bar{\rho}_{12}^{(1)} + Y_2 \mu_m \bar{\rho}_{12}^{(2)} \right) e^{i\omega t} \right]. \tag{4}$$

Similarly, the magnetic field felt by MD2 can be written as

$$B_2(t) = \frac{B_0}{2}e^{-i\omega t} + \frac{B_0}{2}e^{i\omega t} + X_2\mu_0^{-1}\alpha_M\left[\left(\frac{B_0}{2}+Y_1\mu_m\bar{\rho}_{21}^{(1)}+Y_2\mu_m\bar{\rho}_{21}^{(2)}\right)e^{-i\omega t}\right.$$
$$\left.+\left(\frac{B_0}{2}+Y_1\mu_m\bar{\rho}_{12}^{(1)}+Y_2\mu_m\bar{\rho}_{12}^{(2)}\right)e^{i\omega t}\right]. \quad (5)$$

The above matrix elements should satisfy the master equation,

$$\frac{d\rho_{ij}}{dt} = \frac{i}{\hbar}[\rho,H]_{ij} - \Gamma_{ij}\rho_{ij}. \quad (6)$$

To obtain the emission and absorption (extinction) properties of the system, one should know the parameters related to the cavity, namely $\alpha_M$, $X_1$, $X_2$, $Y_1$ and $Y_2$. The high-order magnetic polarizability $\alpha_M$ can be generally obtained numerically by the multipole decomposition method [60], where the near fields are calculated by a numerical finite-difference in time-domain (FDTD) method (Lumerical FDTD). Then, the coefficient $X_1$ can be obtained based on the ratio of the magnetic near field and the $\alpha_M$. $X_2$ can be calculated similarly, while one should note that the two MDs are coherently coupled to the cavity. Thus, there is a phase difference between them. In this work, we focus on a cavity with WGMs. Then, the $X_2$ can be written as $X_2 = X_1 A \cos(\pi\theta/\theta_m)$, where $\theta$ and $\theta_m$ are the angle between the two MDs on the disk circle and the angle between two field nodes (or antinodes) of a WGM mode, respectively. For the interaction between a magnetic dipole and WGM, $Y_1$ ($Y_2$) should be equal to $X_1$ ($X_2$). Similar results have been reported previously [31,33] and here more calculations are carried out to further confirm their relation (see Fig. S1 in the Supplemental Material [61]),. $A$ represents the ratio between the magnetic fields felt by the MDs in the WGM mode. Due to the symmetric reason, the ratio $A$ in a WGM disk cavity is only dependent on the radial locations of the two MDs.

## III. COOPERATIVE EMISSION OF MAGNETIC DIPOLE EMITTERS IN A CAVITY

Let us consider the emission properties of the emitters in the hybrid system as shown in Fig. 1(a). The TE WGM can be excited by MD emitters in the dielectric Si disk. It is known that the power emitted by a MD emitter is $P^{(m)} = \frac{\omega_0}{2}\text{Im}[\mathbf{m}^*\cdot\mathbf{B}]$,

where **m** is the magnetic dipole moment, **B** is the magnetic field felt by the MD emitter, and $\omega_0$ is the oscillating frequency of the MD emitter. The decay rate of the MD emitter $\gamma^{(m)}$ can be written as $\gamma^{(m)} = P^{(m)}/\hbar\omega_0 = \frac{1}{2\hbar}\text{Im}\left[\mathbf{m}^* \cdot \mathbf{B}\right]$. For a MD emitter in vacuum, it is well known that the power radiated by it equals $P_0^{(m)} = \frac{\omega_0^4}{12\pi c^3}\mu_0|\mathbf{m}|^2$, where $c$ is the light speed. Thus, the vacuum decay rate is $\gamma_0^{(m)} = \frac{P_0^{(m)}}{\hbar\omega_0} = \frac{\omega_0^3}{12\pi\hbar c^3}\mu_0|\mathbf{m}|^2$ [15]. The Purcell factor of a MD emitter is defined by $F_p^{(m)} = \frac{P^{(m)}}{P_0^{(m)}} = \frac{\gamma^{(m)}}{\gamma_0^{(m)}} = \frac{\omega_0}{2}\text{Im}\left[\mathbf{m}^* \cdot \mathbf{B}\right]/\frac{\omega_0^4}{12\pi c^3}\mu_0|\mathbf{m}|^2$. For two incoherent MD emitters, the total radiative decay rate in vacuum is $2\gamma_0^{(m)}$. With the chosen disk size, the TE WGMs of $m = 7$, 6 and 5 can be excited in the Si disk at $\lambda = 1235$ nm, 1370 nm and 1546 nm, respectively, as shown in Fig. 1(b). The above working wavelength range is chosen based on the fact that Si is lossless and it has a refractive index of $n \approx 3.5$. The magnetic field distribution ($|B/B_0|$) pattern on the x-y plane of the $m = 6$ mode is shown in Fig. 1(a). Note that this field consists of only the z component, namely, $|B_z/B_0| = |B/B_0|$. Here, we use the ratio $\gamma_{tot}^{(m)}/2\gamma_0^{(m)}$ to characterize the modification of the total decay rate of the two MD emitters due to their cooperative interactions with the dielectric cavity, where $\gamma_{tot}^{(m)}$ is the total decay rate of the two MD emitters in a Si cavity.

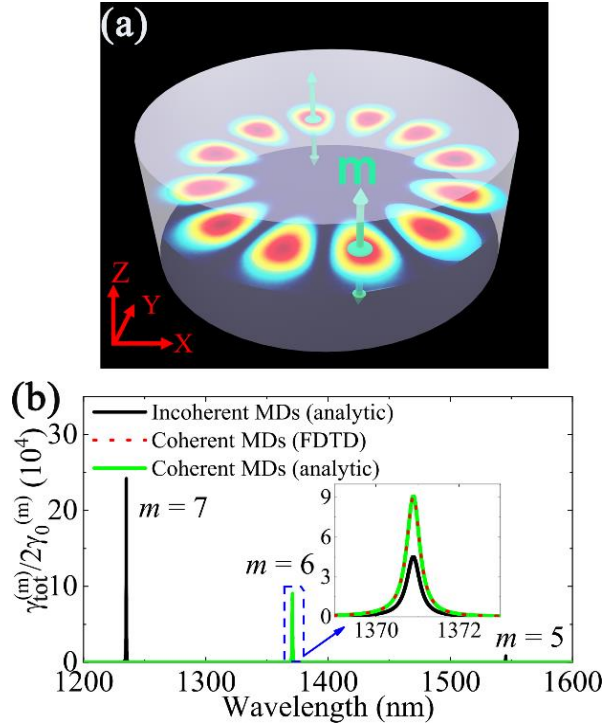

**FIG. 1**. (a) Schematic of a hybrid structure composed of a Si disk cavity and two identical MD emitters. The height and diameter of the disk are $h = 630\ nm$ and $D = 1260\ nm$, respectively. The polarization directions of the MD emitters are both along the $z$-axis. The center of the Si disk is located at (0,0,0). A magnetic field distribution ($|B/B_0|$) pattern on the $x$-$y$ plane of a TE WGM ($m = 6$) is also shown. (b) The total decay rate enhancement $\gamma_{tot}^{(m)}/2\gamma_0^{(m)}$ of the coherent and incoherent MD emitters in the disk cavity. The red dot line and green solid lines are the results obtained from the FDTD and the analytic method, respectively.

We assume that the magnetic field felt by each emitter is the same for simplicity. The ratio $\gamma_{tot}^{(m)}/2\gamma_0^{(m)}$ can be rewritten as

$$\frac{\gamma_{tot}^{(m)}}{2\gamma_0^{(m)}} = \frac{2\frac{1}{2\hbar}\operatorname{Im}\left[\mathbf{m}_{MD}^* \cdot \mathbf{B}_{MD}\right]}{2\frac{\omega_0^3}{12\pi\hbar c^3}\mu_0|\mathbf{m}_{MD}|^2}. \qquad (7)$$

The ratio $\gamma_{tot}^{(m)}/2\gamma_0^{(m)}$ is equal to the Purcell factor of a single MD emitter [15]. This decay rate enhancement in the dielectric Si cavity can be obtained based on our analytical method above. One can rewrite the $\gamma_{tot}^{(m)}/2\gamma_0^{(m)}$ as (see APPENDIX A)

$$\frac{\gamma_{tot}^{(m)}}{2\gamma_0^{(m)}} = \frac{6\pi c^3}{\omega_0^3}\mu_0^{-1}|Y_1|\left|\frac{B}{B_0}\right|, \qquad (8)$$

where $B/B_0$ is the magnetic field enhancement at the point of the MD emitter provided by a WGM in the Si disk, which can be obtained by FDTD simulations [36]. Here the decay rate can be written in the above expression under the condition that the coupling between an emitter and cavity is in a weak coupling regime, and the Femi's golden law applies [9,62]. In this regime, the emitter-cavity coupling strength is not strong enough to surpass the cavity decay rate, which is the case for a common hybrid.

For the coherently interacted MD emitters, the total decay rate can be expressed as $\gamma_{tot}^{(m)} = \frac{1}{2\hbar} \text{Im}\left[\mathbf{m}_{MD}^* \cdot (\mathbf{B}_1 + \mathbf{B}_2)\right]$. The ratio $\gamma_{tot}^{(m)} / 2\gamma_0^{(m)}$ can be rewritten as

$$\frac{\gamma_{tot}^{(m)}}{2\gamma_0^{(m)}} = \frac{\frac{1}{2\hbar} \text{Im}\left[\mathbf{m}_{MD}^* \cdot (\mathbf{B}_1 + \mathbf{B}_2)\right]}{2 \frac{\omega_0^3}{12\pi\hbar c^3} \mu_0 |\mathbf{m}_{MD}|^2}. \tag{9}$$

Based on Eqs. (4) and (5), $\mathbf{B}_2$ can be written as $\mathbf{B}_2 = \mathbf{B}_1 A \cos(\pi\theta/\theta_m)$. For simplicity, we first consider a special configuration where the two MDs are placed at two sides of the disk [Fig. 1(a)]. Due to the symmetry reason, the total magnetic field felt by the each MD is equal to $2\mathbf{B}_1$ or 0 if the two MDs are in phase or out of phase, respectively. For example, the emitters are in phase at the $m = 6$ WGM. The ratio $\gamma_{tot}^{(m)} / 2\gamma_0^{(m)}$ can be expressed as (see APPENDIX B)

$$\frac{\gamma_{tot}^{(m)}}{2\gamma_0^{(m)}} = \frac{12\pi c^3}{\omega_0^3} \mu_0^{-1} |Y_1| \left|\frac{B}{B_0}\right|, \tag{10}$$

which is twice of the incoherent case [Eq. (8)]. The decay rate enhancements $\gamma_{tot}^{(m)} / 2\gamma_0^{(m)}$ for the coherent and incoherent MDs are 90000 and 45000, respectively [Fig. 1(b)]. Note that in Fig. 1(b), the analytic results (solid lines) are obtained by two steps. First, the WGM field profiles ($|B/B_0|$) of a disk are calculated by FDTD simulations with a plane wave as the excitation source (see Fig. S1 in the Supplemental Material [61]). Then, the field enhancement value $|B/B_0|$ is utilized to calculate the decay rate enhancements for different cases based on Eqs. (8-10). The direct FDTD results (the red dotted line) are obtained based on the calculations of the total emitted power of the excitation emitters in a system consisting of a disk and two emitters as the coherent sources.

On the contrary, there is a relative $\pi$ phase change between the magnetic fields produced by MD1 and MD2 at the WGMs of $m = 5$ ($\lambda = 1550$ nm) and $m = 7$ ($\lambda = 1235$ nm), namely $B_1 = -B_2$. This means that the total magnetic field felt by the two coherent MDs is now 0. Therefore, the value of $\gamma_{tot}^{(m)}/2\gamma_0^{(m)}$ is 0 at both the $m = 5$ and $m = 7$ mode. The above cases ($m = 5$ and $m = 7$) are also called subradiant states. A subradiant state is also a characteristic of a system with Dicke effects. In such a state, the radiated fields of two emitters are out of phase with each other and cause a cancellation on their radiations. In our case ($\mathbf{B}_2 = -\mathbf{B}_1$), the radiation of each emitter is totally canceled by the other one. The above results can also be verified by direct FDTD simulations. The results from the two methods show perfect agreement. Here, it should be noted that in the FDTD method, the direct calculated decay rate enhancement value is not $\gamma^{(m)}/\gamma_0^{(m)}$ but $\gamma^{(m)}/\gamma_1^{(m)}$, where $\gamma_1^{(m)} = n^3 \gamma_0^{(m)}$. The details will be discussed later in section **IV**. An electric dipole emitter can also excite the same WGM modes and the decay rate enhancements are relatively much lower. For example, the decay rate enhancement of the $m = 6$ WGM is $\sim 3 \times 10^3$. The cooperative effects are similar to the magnetic countparts (see Fig. S2 in the Supplemental Material [61]), and we shall still focus on the magnetic coherences.

In Fig. 2(a), we study a case where two MD emitters are moving on a circle with a radius of 480 nm. This radius corresponds to the maximal magnetic field enhancement of the $m = 6$ WGM. If the emitters are moving symmetrically on the disk as shown in Fig. 2(a), the magnetic field distribution pattern excited by MDs on the Si disk remains unchanged (see Fig. S3 in the Supplemental Material [61]). Figure 2(b) shows the $\gamma_{tot}^{(m)}/2\gamma_0^{(m)}$ of coherent and incoherent MD emitters as a function of $\theta$. For the incoherent case, each emitter has a decay rate enhancement which does not change with $\theta$ due to the symmetry reason. Therefore, the $\gamma_{tot}^{(m)}/2\gamma_0^{(m)}$ of incoherent MD emitters keeps unchanged at $F_p = 45000$ with varying $\theta$. The red solid line in Fig. 2(b) represents the ratio $\gamma_{tot}^{(m)}/2\gamma_0^{(m)}$ of two coherent MD emitters obtained based on the analytical method. The blue dots are the results carried out by the direct FDTD simulations, which agree well with that calculated based on the analytic method. $\gamma_{tot}^{(m)}/2\gamma_0^{(m)}$ varies as a trigonometric function of $\theta$. This is caused by the fact that the relative phase difference between the fields produced by $m_1$ and $m_2$ is $\cos(\pi\theta/\theta_m)$. Note that the $\gamma_{tot}^{(m)}/2\gamma_0^{(m)}$ does not decay with the distance between the emitters. Similar distance dependence behaviors have been reported in many high-$Q$ WGM related phenomena, for example, Rayleigh's observation sound of WGM. This is due to the facts that the cavity does not have material loss and its radiative loss is very low with such a high $Q$-factor. This is a significant advantage over the plasmonic system where the cooperative effect suffers from high losses[10,57].

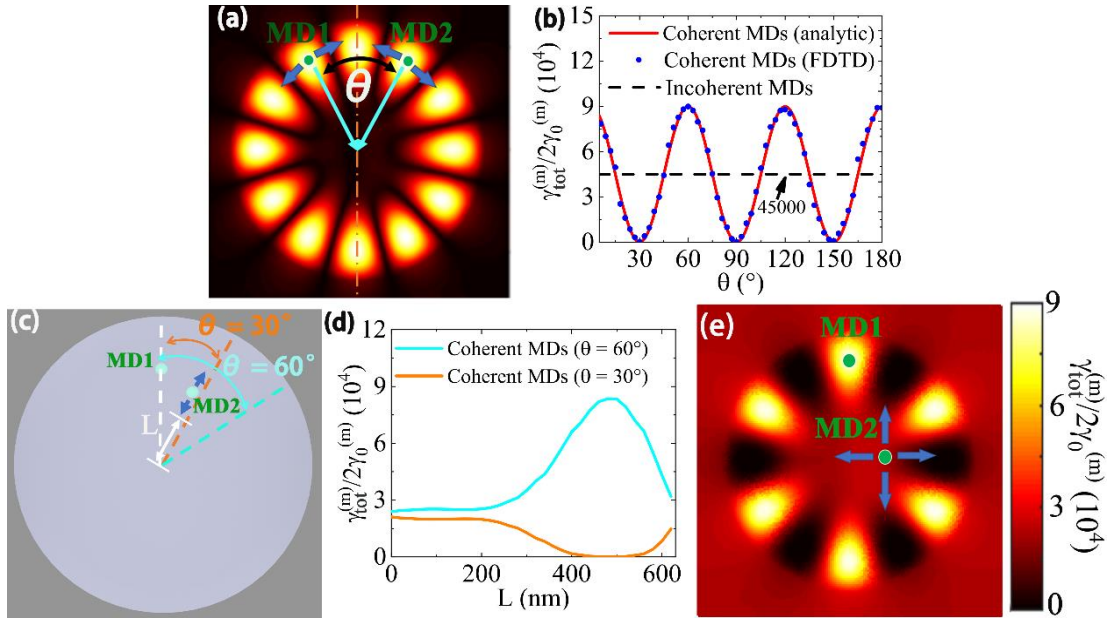

**FIG. 2**. (a) Magnetic field enhancement distribution ($|B/B_0|$) of a TE WGM ($m = 6$) in the Si disk. The MD emitters move on a circle with a radius of 480 nm, and the distance between the two emitters is denoted by the angle $\theta$. (b) The ratio $\gamma_{tot}^{(m)}/2\gamma_0^{(m)}$ of the coherent and incoherent MD

emitters as a function of $\theta$. (c) The schematic of the location of the two coherent MD emitters with different radial locations. The location of the MD1 is fixed. MD2 can move on the radial lines of $\theta = 30°$ or $\theta = 60°$. The distance between MD2 and the disk center is $L$. (d) The ratio $\gamma_{tot}^{(m)}/2\gamma_0^{(m)}$ of coherent MD emitters as a function of $L$. (e) The distribution of the ratio $\gamma_{tot}^{(m)}/2\gamma_0^{(m)}$ for two coherent MD emitters, where the MD1 is fixed and MD2 moves on the whole plane.

There are challenges to accurately control the radial location of the MD emitters in experiments. Therefore, it is also important to study the cooperative effects between the two MD emitters at locations with different field enhancements inside the cavity. For simplicity, we assume that the location of MD1 is fixed at (0, 480, 0) nm, and MD2 moves on a line of $\theta = 30°$ or $\theta = 60°$. The distance between MD2 and the center of the Si disk is $L$. The ratios $\gamma_{tot}^{(m)}/2\gamma_0^{(m)}$ of coherent two MDs are shown in Fig. 2(d). Due to the relative π-phase difference between the magnetic fields felt by MDs for the condition of $\theta = 30°$, the total magnetic field felt by each MD becomes lower and eventually reach to 0 when MD2 moves to the location ($L = 480$ nm) where its magnetic field enhancement is equal to that of MD1. Thus, the coherent MDs undergo destructive cooperative effects and the ratio $\gamma_{tot}^{(m)}/2\gamma_0^{(m)}$ approaches 0 around $L = 480$ nm. On the contrary, the magnetic fields felt by MD1 and MD2 are in phase when MD2 is moving on the line of $\theta = 60°$, which cause constructive effects between MDs and the ratio $\gamma_{tot}^{(m)}/2\gamma_0^{(m)}$ reaches the maximum (~90000) at $L= 480$ nm [Fig. 2(d)]. Thus, the cooperative emission rate is highly related to the field distribution pattern of a WGM. Figure 2(e) shows a case where MD1 is still fixed at the location (0, 480 nm, 0) and MD2 moves in all directions on the whole $x$-$y$ plane with the $m = 6$ mode. The value represents the ratio $\gamma_{tot}^{(m)}/2\gamma_0^{(m)}$ of the two MDs when the MD2 is moved to the corresponding location. The $\gamma_{tot}^{(m)}/2\gamma_0^{(m)}$ distribution pattern is similar to the magnetic field distribution pattern of the $m = 6$ WGM.

Based on the discussion of two MDs, the cooperative effects in such kind of system can also be expanded to hybrid system with $N$ coherent emitters. The total decay rate of the $N$ coherent MDs can be expressed as (see APPENDIX B)

$$\gamma_{tot}^{(m)} = \frac{(X_1 + X_2 + \cdots\cdots + X_N)(Y_1 + Y_2 \cdots\cdots + Y_N)}{X_1 Y_1} \frac{\omega}{2\hbar\omega_0} \text{Im}\left[\mathbf{m}_{MD}^* \cdot \mathbf{B}_1\right], \tag{11}$$

where $X_N$ and $Y_N$ are the proportional coefficients of the MD emitter $N$. The relative ratio between $X_N$ ($Y_N$) of MD emitter $N$ and $X_1$ ($Y_1$) of MD1 is $A_N \cos(\pi \theta_N/\theta_m)$, where $\theta_N$ is the angle difference between emitter $N$ and MD1, and $A_N$ is the magnitude ratio. Figure 3(a) shows a case of three emitters located on the $x$-$y$ plane. For simplicity, the distance between each MD and the disk center is the same ($L = 480$ nm). The working wavelength is still around $\lambda = 1370$ nm corresponding to the WGM of $m = 6$. Thus, the ratio $X_3/X_1$ now reduces to $\cos(\pi\theta_3/30°)$. The $\theta_2$ and $\theta_3$ are 30° and 90°, respectively. Both MD2 ($\pi\theta_2/30° = \pi$) and MD3 ($\pi\theta_3/30° = 3\pi$) are out of phase with MD1, respectively. The ratio $\gamma_{tot}^{(m)}/3\gamma_0^{(m)}$ of the three MD emitters reaches a peak value of 15000 which is only 1/3 of a single MD emitter [Fig. 3(a)]. By fixing the locations of MD1 and MD2, the $\gamma_{tot}^{(m)}/3\gamma_0^{(m)}$ distribution with moving the MD3 on the $x$-$y$ plane is shown in Fig. 3(b). The field distribution pattern is similar to the case with only a single emitter as the excitation source. However, the total decay rate enhancement of the three MDs becomes much lower. This is due to the fact that the MD1 and MD2 are still canceled with each other, namely, $X_1+X_2 = 0$ and $Y_1+Y_2 = 0$. The $\gamma_{tot}^{(m)}$ reduces to $X_3 Y_3/X_1 Y_1 \propto (A_N)^2$. Thus, the maximal $\gamma_{tot}^{(m)}/3\gamma_0^{(m)}$ is only 1/3 of that of a single MD emitter.

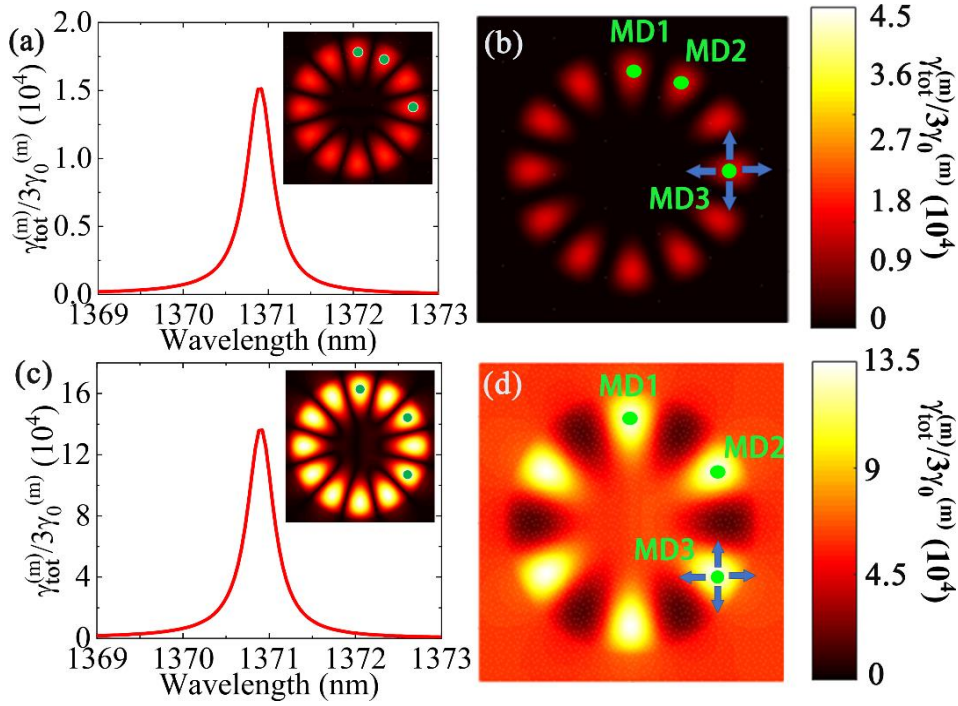

**FIG. 3.** (a) The ratio $\gamma_{tot}^{(m)}/3\gamma_0^{(m)}$ of three MDs. The locations of the three MDs with the corresponding field distribution are shown in the inset. The WGM cavity is the same as that in Fig.1. (b) The ratio $\gamma_{tot}^{(m)}/3\gamma_0^{(m)}$ distribution of three coherent MD emitters. The locations of MD1 and MD2 are fixed while MD3 moves on the plane. (c) and (d) show the same contents as (a) and (b), respectively, while the MD2 is fixed at a different location.

We also consider another case where the locations of MD1 and MD2 are still fixed, while they are in phase. This can be realized by setting MD2 at $\theta_2 = 60°$ for the $m = 6$ WGM [Fig. 3(c)]. By moving the MD3 on the *x-y* plane, the maximal value of the ratio $\gamma_{tot}^{(m)}/3\gamma_0^{(m)}$ of the emitters in the hybrid system can reach 135000 when MD3 is in phase with both MD1 and MD2. This value is three times larger than the incoherent case [Figs. 3(c) and 3(d)]. Note that the field distributions are highly dependent on the emitter locations. The total electromagnetic fields excited by the three MDs in Si disk cavity are a coherent superposition of that excited by each one (see Fig. S4 in the Supplemental Material [61]).

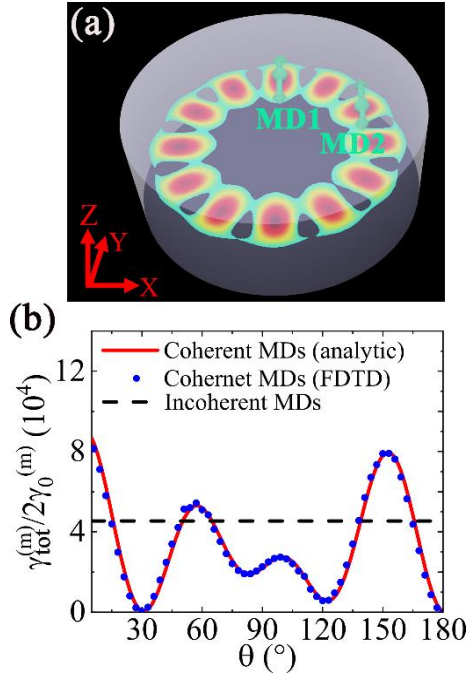

**FIG. 4.** (a) Schematic of a hybrid system consisting of a TM WGM ($m = 6$) cavity and two coherent MD emitters. (b) The ratio $\gamma_{tot}^{(m)}/2\gamma_0^{(m)}$ of the coherent and incoherent MD emitters as a function of $\theta$.

In the above investigations, the polarizations of the emitters are always along the directions of magnetic fields of a TE WGM. Here, we also consider another experimentally realistic case where the polarization of each emitter is always along a given direction and the WGM is a TM mode. This means that there are certain angles between the directions of emitters and the magnetic fields of a WGM. Thus, even though a MD is moving on a circle with a fixed radius, its decay rate enhancement still varies with its location. This is the main difference compared to the above cases where the decay rate enhancement of a MD does not vary with location on a given circle. Figure 4(a) shows a case where the polarization of both MD emitters is always along the $y$-axis. The WGM is a TM resonance with $m = 6$ of the same Si disk as before. The resonance wavelength is $\lambda = 1502$ nm. For simplicity, MD1 is fixed at a location with $r = 460$ nm. The magnetic field felt by it $B_1$ can now be expressed as

$B_1 = X_1 m + X_2 m \cos\theta \cos\frac{180°}{30°}\theta$. The electric field felt by MD2 can be expressed as $B_2 = X_2 m \cos\theta + X_1 m \cos\frac{180°}{30°}\theta$. Therefore, the ratio $\gamma_{tot}^{(m)}/2\gamma_0^{(m)}$ becomes

$$\frac{\gamma_{tot}^{(m)}}{2\gamma_0^{(m)}} = \frac{1+\cos\theta\cos6\theta+(\cos\theta+\cos6\theta)\cos\theta}{2}\frac{6\pi c^3 \omega}{\omega_0^4}\mu_0^{-1}|Y_1|\left|\frac{B}{B_0}\right|. \qquad (12)$$

The results are plotted in Fig. 4(b), and they agree well with the direct numerical FDTD simulation.

## IV. EMISSION OF AN EMITTER WITH AND WITHOUT A SMALL HOLE IN A CAVITY

It should be noted that the $\gamma^{(m)}/\gamma_0^{(m)}$ of a MD in the dielectric cavity cannot be further enhanced by introducing a small hole in the cavity. One may find that the decay rate enhancements given directly by FDTD simulations for a cavity with and without a small hole are quite different as shown in Figs. 5(a) and 5(b). This is due the fact that the decay rate enhancement expressions for the two cases are different. The expressions for a cavity with and without a small hole are $\gamma^{(m)}/\gamma_0^{(m)}$ and $\gamma^{(m)}/\gamma_1^{(m)}$ respectively. $\gamma_1^{(m)}$ represents the decay rate of a MD in a dielectric medium, and $\gamma_1^{(m)} = n^3 \gamma_0^{(m)}$ [15]. Considering this fact, the $\gamma^{(m)}/\gamma_0^{(m)}$ by the two simulations give the same results [Fig. 5(b)]. This can also be further confirmed by the corresponding magnetic field enhancement ($|B/B_0|$) of the resonance mode, which is highly related to the $\gamma^{(m)}/\gamma_0^{(m)}$ (see Eq. (8)). $|B/B_0|$ will not be further enhanced by introducing a small hole as shown in Fig. 5(a). This is due to the fact that the hole is along the direction of the magnetic field. Furthermore, the relative permeability of the common dielectric material is 1. Thus, both the $|B/B_0|$ and the corresponding $\gamma^{(m)}/\gamma_0^{(m)}$ are still not further enhanced for a TM case (see Fig. S5 in the Supplemental Material [61]). This feature may be broken for the electric counterpart as the dielectric constant of the Si is larger than 1 here. Figures 5(c) and 5(d) show a case with the same WGM while the emitter is replaced by an electric dipole. For an electric dipole, its decay rate in a medium of index $n$ ($\gamma_1^{(p)}$) is $\gamma_1^{(p)} = n\gamma_0^{(p)}$, where $\gamma_0^{(p)}$ is the decay rate in vacuum. In the direct FDTD simulations, the decay rate enhancements with and without a small hole are $\gamma^{(p)}/\gamma_0^{(p)}$ and $\gamma^{(p)}/\gamma_1^{(p)}$ respectively. The electric field in the hole can be further enhanced due to the continuity of electric displacement vector [Fig. 5(c)] [35]. The decay rate enhancement $\gamma^{(p)}/\gamma_0^{(p)}$, which is highly related to the electric field enhancement, also increases [Fig. 5(d)]. On the other hand, if the direction of electric displacement vector is oriented along the hole (z-axis) with a TM WGM mode, the existence of the gap does not further increase the electric field enhancement [Fig. 5(e)]. Then, the $\gamma^{(p)}/\gamma_0^{(p)}$ is not further enhanced in the small hole [Fig. 5(f)].

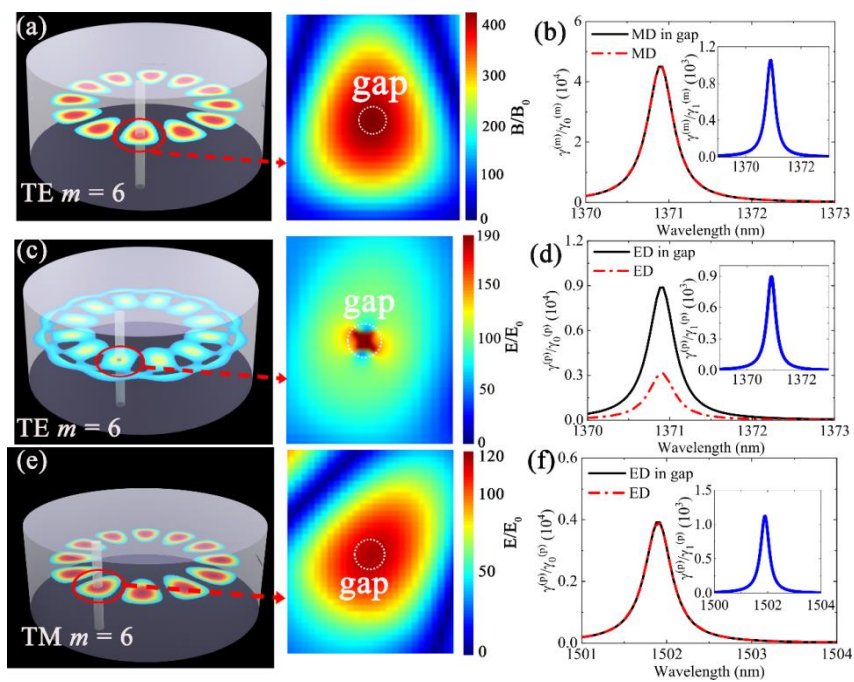

**FIG. 5.** (a) The magnetic field distribution of the TE $m = 6$ WGM on the *x-y* plane in the Si disk. A hole with a radius of 20 nm is introduced at the location of a field antinode. (b) The ratios of $\gamma^{(m)}/\gamma_0^{(m)}$ in a solid disk and a disk with a small hole. The inset shows the ratio $\gamma^{(m)}/\gamma_1^{(m)}$ of the MD in a solid disk that calculated directly by the FDTD. (c) and (d) show the same contexts of the (a) and (b), respectively, while the emitter is replaced by an ED one. The corresponding expressions for decay rate ratios of an ED are $\gamma^{(p)}/\gamma_0^{(p)}$ and $\gamma^{(p)}/\gamma_1^{(p)}$. (e) and (f) show the same contexts of the (c) and (d), respectively, while the working mode is replaced by a TM WGM of $m = 6$.

## V. ABSORPTION OF COHERENT MAGNETIC DIPOLE EMITTERS IN A CAVITY

We have shown that the decay rate of the emitters can be greatly affected by their coherent couplings. We shall now consider the influence of the coherent coupling between MDs on the absorption (or extinction) properties of the hybrid system. Our results show that the coupling strength between each emitter and the cavity is hardly affected by the other coherent emitters. Figure 6(a) shows a dielectric Si disk and two emitters excited by a plane wave with polarization in the *y*-axis. A TE WGM of $m = 5$ can be excited at $\lambda = 1545$ nm. Two identical magnetic emitters are placed at the locations with the maximal |B/B$_0$|. The treatment is different from the emission based on the fact that the magnetic field of the excitation plane wave (B$_0$) should be taken into account. By solving the Eq. (6) in the steady-state condition, one can obtain the absorption properties. In fact, the situation here is quite similar to our previous work where the emitters are placed in a same near-field antinode of a WGM [33]. The emitters were taken as a single equivalent emitter there, while the emitters here are considered as separated ones. If the field felt by the emitters in this work is the same, then the two treatments become the same. Here, for the sake of completeness, we have included the expressions in the Supplemental material [61]. For simplicity, we assume that relaxation elements of diagonal (the longitudinal homogeneous lifetime) and nondiagonal elements (the transverse homogeneous lifetime) are the same, which correspond to $\Gamma_{ij} = 1/T$. The total extinction cross section $(\sigma_{ext})_{MD}$ of the two MD emitters can be expressed as

$$(\sigma_{ext})_{MD} = \frac{4\pi}{\lambda} \frac{\mu_0 \mu_m^2}{\hbar} \frac{T}{1 + (\omega - \omega_0)^2 T^2 + \frac{\mu_m^2}{\hbar^2} |B_{MD}|^2 T^2} \frac{|B_{MD}|^2}{B_0^2}. \tag{13}$$

The extinction section of the Si disk resonator can be divided into two parts. One is the contribution from a WGM of the Si disk resonator, which is coupled with the MDs, and the other one is the background contribution from other broad-band modes which is assumed to be unaffected by MDs. Therefore, the extinction cross section of the coupled Si disk cavity can be written as

$$(\sigma_{ext})_{cav} = \frac{2\pi}{\lambda} \text{Im}(\alpha_M) \frac{|B_{cav}|^2}{B_0^2} + C, \tag{14}$$

where $C$ represents the contribution of the other unaffected modes.

The intensity of the excitation plane wave is $I_0 = 10$ W/cm$^2$, which satisfies $\mu_m^2/\hbar^2 |B_{tot}| T^2 \ll 1$. The responses of the MDs and the disk cavity are in linear region. We take $T$ to be 4 ns. For clearer illustration, the $\mu_m$ is taken to be a large value $\mu_m = 20\,\mu_B$ ($\mu_B$ is the Bohr magneton). The spectral width of the MDs is much narrower than that of the WGM. Figures 6(c) and 6(d) shows the extinction and absorption spectra of the hybrid system obtained by the above expressions [Eqs. (13) and (14)]. A pronounced dip appears on the spectrum which represents efficient coupling between emitters and the cavity. The corresponding absorption is enhanced ~$10^4$ times. However, the absorption of an emitter is not further enhanced by introducing another coherent emitter. This is further confirmed by moving the MD2 to a neighboring near-field antinode [Fig. 6(b)] whose phase is changed by π. The extinction and absorption spectra are numerically the same as that before [Figs. 6(c) and 6(d)]. The cooperative behavior of the absorption is quite different from the emission. The difference is highly related to the excitation of a WGM. Without losing of generality, let us consider a two emitters case. In the emission process, a WGM is excited by the emitters themselves ($Y_1 m_1 + Y_2 m_2$). Thus, the field felt by an emitter is contributed by itself and the other one. In the absorption case, a WGM is firstly excited by an outside plane wave source ($B_0$) and the contribution from the emitters is much smaller compared to the excitation source. Thus, the field felt by an emitter is mainly contributed from the plane wave-excited WGM, and the contribution from the other coherent emitter is much smaller. This explains why the emission rate of an emitter is greatly affected by another one, while the absorption of the emitter is hardly affected by another one.

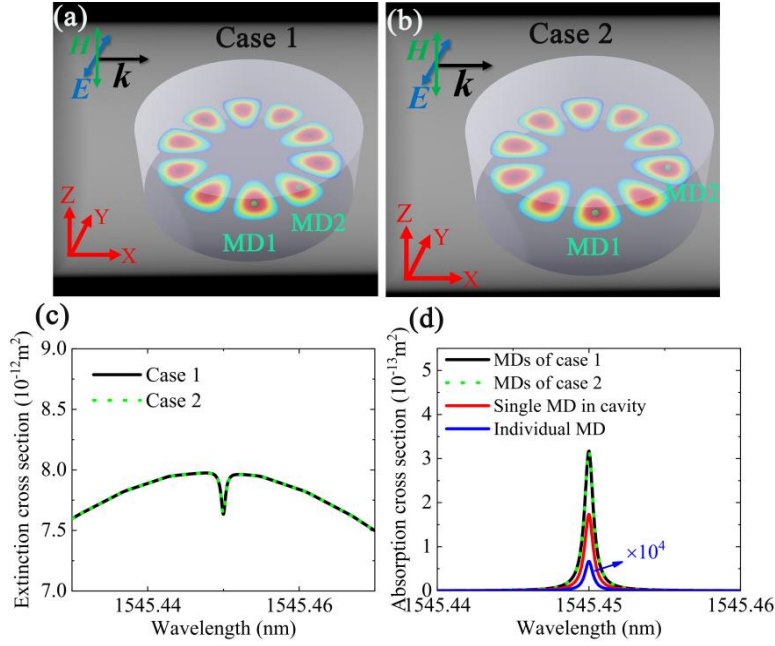

**FIG. 6.** (a, b) Schematic of a hybrid system of (a) case 1 and (b) case 2. The cavity is the same as that in Fig. 1. (c) The extinction spectra of the hybrid system. (d) The absorption spectra of the coherent MDs in the cavity. The individual MD and a single MD in the cavity is also shown for comparison.

## VI. CONCLUSION

In conclusion, we have theoretically investigated the radiation properties of coherently interacting MD emitters mediated by a subwavelength Si disk of high-$Q$ WGMs. The efficient emitter-WGM coupling can induce both a high emission rate enhancement ($\sim 10^4$) of a single emitter and efficient coherences between MD emitters. In a WGM cavity, the contribution to the emission of an emitter from another one is highly dependent on the relative phase difference and the magnetic field enhancement provided by the WGM. When one MD emitter is placed at a near-field antinode of a WGM, a maximal emission rate ($\gamma_{max}$) occurs for a single emitter. Then, the maximal contribution from another coherent MD emitter can also reach $\gamma_{max}$ Thus, the emission of an MD emitter can be largely further enhanced with many other coherent emitters. Importantly, the cooperative emission rate of the coherent MD emitters does not decay with distance in our cavity due to the high-$Q$ feature. This is a significant advantage over the plasmonic system which suffers from high losses. Note that the decay rate of a MD in the dielectric cavity cannot be further enhanced by introducing a small hole in the cavity. We also consider absorption properties of the emitters in such a hybrid system. However, the absorption of an emitter is hardly affected by the coherent couplings with other emitters. The difference between the cooperative performances of the emission and absorption is highly related to the fact that the WGMs involved are excited differently. The strong magnetic superradiance and the related efficient coherences are important for enhanced magnetic light-matter interactions.

**ACKNOWLEDGMENTS**

This paper was supported by the National Natural Science Foundation of China (No. 11704416), the Hunan Provincial Natural Science Foundation of China (No. 2021JJ20076).

**APPENDIX A: Purcell factor of a single MD emitter in a cavity**

For an MD emitter with an oscillating MD moment $\mathbf{m}$, the decay rate in free space can be expressed as

$$\gamma_0^{(m)} = \frac{\omega_0^3}{12\pi\hbar c^3}\mu_0|\mathbf{m}|^2, \tag{A1}$$

where $\omega_0$ is the frequency of the spontaneous emission of MD emitter, $\hbar$ is the reduced Planck constant. $c$ is the speed of light, $\mu_0$ is the vacuum permeability. For the MD placed in resonant nanocavities, the decay rate can be expressed as

$$\gamma^{(m)} = \frac{1}{2\hbar}\text{Im}\left[\mathbf{m}^* \cdot \mathbf{B}_{MD}\right], \tag{A2}$$

where $\mathbf{B}_{MD}$ is the magnetic field felt by the MD emitter. The magnetic field felt by the cavity can be written as $B_{cav} = Y_1\mathbf{m}$, therefore the magnetic field felt by the MD emitter can be expressed as $\mathbf{B}_{MD} = X_1 m_{cav} = X_1\mu_0^{-1}\alpha_M B_{cav} = X_1 Y_1 \mu_0^{-1}\alpha_M \mathbf{m}$, where

$m_{cav}$ is the high-order magnetic moment of the nanocavity, which can be expressed as $m_{cav} = \mu_0^{-1} \alpha_M B_{cav}$, $\alpha_M$ is the high-order magnetic polarizability of the cavity. Now the decay rate $\gamma^{(m)}$ can be rewritten as

$$\begin{aligned} \gamma^{(m)} &= \frac{1}{2\hbar} \text{Im}\left[\mathbf{m} \cdot X_1 Y_1 \mu_0^{-1} \alpha_M \mathbf{m}\right] \\ &= \frac{1}{2\hbar} \text{Im}\left[Y_1 \frac{B}{B_0} \mathbf{m}^2\right], \end{aligned} \quad (A3)$$

where $B/B_0 = X_1 \mu_0^{-1} \alpha_M$ is the magnetic field enhancement in the position of the MD emitter, which is produced by the nanocavity excited by a plane wave with the magnetic field $B_0$. Thus, the Purcell factor can be written as

$$\begin{aligned} F_p^{(m)} = \gamma^{(m)}/\gamma_0^{(m)} &= \frac{1}{2\hbar} \text{Im}\left[Y_1 \frac{B}{B_0} \mathbf{m}^2\right] \Big/ \frac{\omega_0^3}{12\pi\hbar c^3} \mu_0 |\mathbf{m}|^2 \\ &= \frac{6\pi c^3}{\omega_0^3} \mu_0^{-1} \text{Im}\left[Y_1 \frac{B}{B_0}\right] \\ &= \frac{6\pi c^3}{\omega_0^3} \mu_0^{-1} |Y_1| \left|\frac{B}{B_0}\right|. \end{aligned} \quad (A4)$$

For the two incoherent MD emitters in cavities, the ratio $\gamma_{tot}^{(m)}/2\gamma_0^{(m)}$ is equal to the Purcell factor of a single emitter above.

**APPENDIX B: Emission properties of coherent MD emitters in a cavity**

We now consider the case of placing the two coherent MD emitters in a cavity. We assume that the two MD emitters are identical. In this case, the magnetic field felt by the cavity can be expressed as $B_{cav} = (Y_1 + Y_2)\mathbf{m}$, and the high-order magnetic moment of the cavity is $m_{cav} = \mu_0^{-1} \alpha_M B_{cav} = \mu_0^{-1} \alpha_M (Y_1 + Y_2)\mathbf{m}$. The magnetic field felt by MD emitters can be written as $B_1 = X_1 \mu_0^{-1} \alpha_M (Y_1 + Y_2)\mathbf{m}$ and $B_2 = X_2 \mu_0^{-1} \alpha_M (Y_1 + Y_2)\mathbf{m}$, respectively. Therefore, the total decay rate $\gamma_{tot}^{(m)}$ of the two MD emitters can be expressed as

$$\gamma_{tot}^{(m)} = \frac{1}{2h}\operatorname{Im}\left[\mathbf{m}^{*}\cdot\mathbf{B}_{1}\right]+\frac{\omega}{2h\omega_{0}}\operatorname{Im}\left[\mathbf{m}^{*}\cdot\mathbf{B}_{2}\right]$$
$$= \frac{1}{2h}\operatorname{Im}\left[(X_{1}+X_{2})(Y_{1}+Y_{2})\mu_{0}^{-1}\alpha_{M}\mathbf{m}^{2}\right].$$
(B1)

The ratio $\gamma_{tot}^{(m)}/2\gamma_{0}^{(m)}$ can be expressed as

$$\frac{\gamma_{tot}^{(m)}}{2\gamma_{0}^{(m)}} = \frac{1}{2h}\operatorname{Im}\left[(X_{1}+X_{2})(Y_{1}+Y_{2})\mu_{0}^{-1}\alpha_{M}\mathbf{m}^{2}\right]/2\frac{\omega_{0}^{3}}{12\pi\hbar c^{3}}\mu_{0}|\mathbf{m}|^{2}$$
$$= \frac{3\pi c^{3}}{\omega_{0}^{3}}\mu_{0}^{-1}(Y_{1}+Y_{2})\operatorname{Im}\left[(1+\frac{X_{2}}{X_{1}})\frac{B}{B_{0}}\right]$$
$$= \frac{3\pi c^{3}}{\omega_{0}^{3}}\mu_{0}^{-1}(Y_{1}+Y_{2})(1+\frac{X_{2}}{X_{1}})\left|\frac{B}{B_{0}}\right|.$$
(B2)

As for the case of $N$ MD emitters in a cavity, the ratio $\gamma_{tot}^{(m)}/N\gamma_{0}^{(m)}$ can be expressed as

$$\frac{\gamma_{tot}^{(m)}}{N\gamma_{0}^{(m)}} = \frac{1}{2h}\operatorname{Im}\left[(X_{1}+X_{2}+\cdots+X_{N})(Y_{1}+Y_{2}+\cdots+Y_{N})\mu_{0}^{-1}\alpha_{M}\mathbf{m}^{2}\right]/N\frac{\omega_{0}^{3}}{12\pi\hbar c^{3}}\mu_{0}|\mathbf{m}|^{2}$$
$$= \frac{6\pi c^{3}}{N\omega_{0}^{3}}\mu_{0}^{-1}(Y_{1}+Y_{2}+\cdots+Y_{N})\operatorname{Im}\left[(1+\frac{X_{2}}{X_{1}}+\cdots+\frac{X_{N}}{X_{1}})\frac{B}{B_{0}}\right]$$
$$= \frac{6\pi c^{3}}{N\omega_{0}^{3}}\mu_{0}^{-1}(Y_{1}+Y_{2}\cdots+Y_{N})(1+\frac{X_{2}}{X_{1}}\cdots+\frac{X_{N}}{X_{1}})\left|\frac{B}{B_{0}}\right|.$$
(B3)

results of Figs. S1–S5.

Supplemental Material for

# "Strong Superradiance of Coherently Coupled Magnetic Dipole Emitters Mediated by Whispering Gallery Modes of a Subwavelength All-Dielectric Cavity"

Ma-Long Hu, Xiao-Jing Du, Lin Ma, Jun He and Zhong-Jian Yang*

*Hunan Key Laboratory of Nanophotonics and Devices, School of Physics and Electronics, Central South University, Changsha 410083, China*

*E-mail: zjyang@csu.edu.cn

**Part 1: The extinction (absorption) cross sections of the hybrid system.**

The magnetic field felt by MD1 can be written as

$$B_1(t) = \frac{B_0}{2}e^{-i\omega t} + \frac{B_0}{2}e^{i\omega t} + X_1\mu_0^{-1}\alpha_M\left[\left(\frac{B_0}{2} + Y_1\mu_m\bar{\rho}_{21}^{(1)} + Y_2\mu_m\bar{\rho}_{21}^{(2)}\right)e^{-i\omega t} \right. \\ \left. + \left(\frac{B_0}{2} + Y_1\mu_m\bar{\rho}_{12}^{(1)} + Y_2\mu_m\bar{\rho}_{12}^{(2)}\right)e^{i\omega t}\right]. \quad \text{(S1)}$$

Similarly, the magnetic field felt by MD2 can be written as

$$B_2(t) = \frac{B_0}{2}e^{-i\omega t} + \frac{B_0}{2}e^{i\omega t} + X_2\mu_0^{-1}\alpha_M\left[\left(\frac{B_0}{2} + Y_1\mu_m\bar{\rho}_{21}^{(1)} + Y_2\mu_m\bar{\rho}_{21}^{(2)}\right)e^{-i\omega t} \right. \\ \left. + \left(\frac{B_0}{2} + Y_1\mu_m\bar{\rho}_{12}^{(1)} + Y_2\mu_m\bar{\rho}_{12}^{(2)}\right)e^{i\omega t}\right]. \quad \text{(S2)}$$

Now we consider the case where MD1 and MD2 are identical, which means $\rho_{12}^{(1)} = \rho_{12}^{(2)}$, $\rho_{21}^{(1)} = \rho_{21}^{(2)}$, $X_1(Y_1) = X_2(Y_2)$. Thus, the magnetic fields felt by the emitter 1 and emitter 2 can be rewritten as

$$B_1(t) = \frac{\hbar}{\mu_m}\left[\left(\Omega + G\rho_{21}^{(1)}\right)e^{-i\omega t} + \left(\Omega^* + G^*\rho_{12}^{(1)}\right)e^{i\omega t}\right], \tag{S3}$$

where $\Omega = \frac{\mu_m}{\hbar}\left(1 + X_1\mu_0^{-1}\alpha_M\right)\frac{B_0}{2}$, $G = \frac{\mu_m^2}{\hbar}\left(2X_1\mu_0^{-1}\alpha_M Y_1\right)$. Based on the master equation, we have

$$\begin{aligned}
\frac{d\rho_{12}^{(1)}}{dt} &= -i\omega_0\rho_{21}^{(1)} + \frac{i\mu_m B_1(t)}{\hbar}\left(\rho_{11}^{(1)} - \rho_{11}^{(2)}\right) - \frac{\rho_{21}^{(1)}}{T}, \\
\frac{d\left(\rho_{11}^{(1)} - \rho_{22}^{(1)}\right)}{dt} &= \frac{2i\mu_m B_1(t)}{\hbar}\left[\rho_{21}^{(1)} - \left(\rho_{21}^{(1)}\right)^*\right] - \frac{\left(\rho_{11}^{(1)} - \rho_{22}^{(1)}\right) - \left(\rho_{11}^{(1)} - \rho_{22}^{(1)}\right)_0}{T},
\end{aligned} \tag{S4}$$

where $\left(\rho_{11}^{(1)} - \rho_{11}^{(1)}\right)_0$ is the initial population difference of MD1, which is taken to be $\left(\rho_{11}^{(1)} - \rho_{11}^{(1)}\right)_0 = 1$. We write $\Delta = \rho_{11}^{(1)} - \rho_{11}^{(1)}$, $\bar{\rho}_{12} = C + iD, \bar{\rho}_{21} = C - iD$. With the rotating wave approximation, we find

$$\begin{aligned}
\frac{dC}{dt} &= -\frac{C}{T} + (\omega - \omega_0)D - (\Omega_I + G_I C - G_R D)\Delta, \\
\frac{dD}{dt} &= -\frac{D}{T} - (\omega - \omega_0)C - (\Omega_R + G_R C - G_I D)\Delta, \\
\frac{d\Delta}{dt} &= \frac{1-\Delta}{T} + 4\Omega_I C + 4\Omega_R D + 4G_I\left(C^2 + D^2\right),
\end{aligned} \tag{S5}$$

where $\Omega_R$ and $\Omega_I$ are the real and imaginary parts of $\Omega$, respectively. $G_R$ and $G_I$ are the real and imaginary parts of $G$, respectively. In the steady-state situation, the left-hand side of Eq. (S8) satisfies $\frac{dC}{dt} = \frac{dD}{dt} = \frac{d\Delta}{dt} = 0$. We obtain

$$\frac{4\Delta}{T}\frac{\Omega_I^2 + \Omega_R^2}{\left(\frac{1}{T} + G_I\Delta\right)^2 + \left[(\omega - \omega_0) + G_R\Delta\right]^2} = \frac{1-\Delta}{T_1},$$

$$C = \frac{\Delta\Omega_R\left[(\omega - \omega_0) + \Delta G_R\right] + \Delta\Omega_I\left(\frac{1}{T} + G_I\Delta\right)}{\left(\frac{1}{T} + G_I\Delta\right)^2 + \left[(\omega - \omega_0) + G_R\Delta\right]^2}, \tag{S6}$$

$$D = \frac{\Delta\Omega_I\left[(\omega - \omega_0) + \Delta G_R\right] + \Delta\Omega_R\left(\frac{1}{T} + G_I\Delta\right)}{\left(\frac{1}{T} + G_I\Delta\right)^2 + \left[(\omega - \omega_0) + G_R\Delta\right]^2}$$

Based on Eq. (2), the MD moment of MD1 can be written as

$$m_1 = \frac{\mu_m^2}{\hbar} \frac{(\omega - \omega_0)T^2 + iT}{1 + (\omega - \omega_0)^2 T^2 + \frac{\mu_m^2}{\hbar^2}|B_1|^2 T^2} B_1. \tag{S7}$$

The MD moment of MD1 can be also expressed as $m_1 = \mu_0^{-1}\alpha_{MD1}B_1$, where $\alpha_{MD1}$ is the magnetic dipole polarizability of MD1. Thus, it can be rewritten as

$$\alpha_{MD1} = \frac{\mu_0 \mu_m^2}{\hbar} \frac{(\omega - \omega_0)T^2 + iT}{1 + (\omega - \omega_0)^2 T^2 + \frac{\mu_m^2}{\hbar^2}|B_1|^2 T^2}. \tag{S8}$$

Since the two emitters are identical, the properties of MD2 are the same. The extinction (absorption) cross sections of the hybrid system can be calculated based on the above results.

## Part 2: Figures S1-S5.

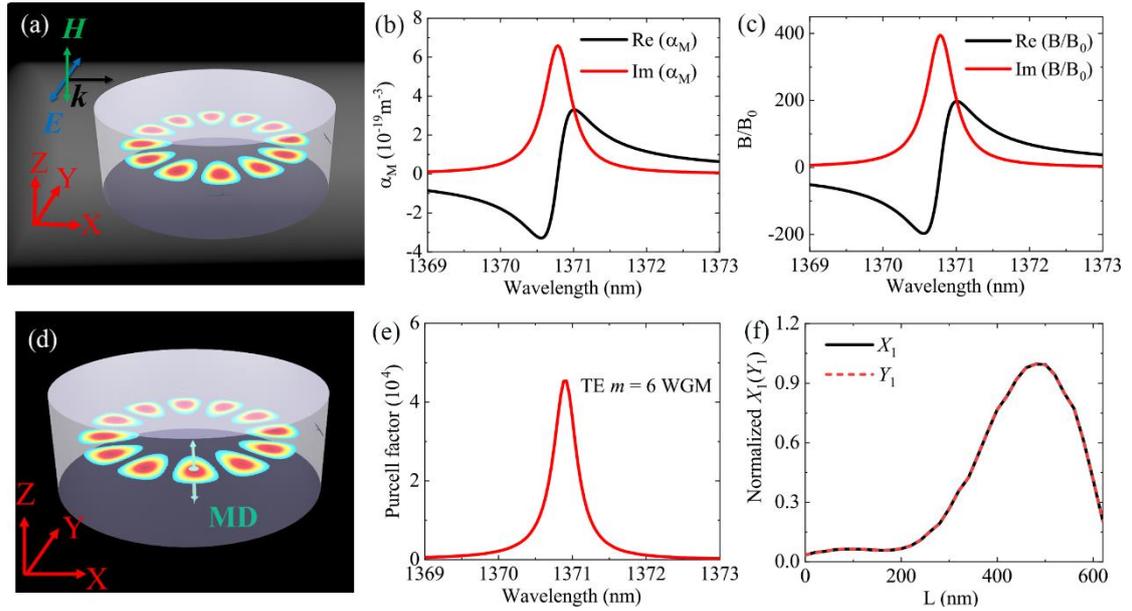

**FIG. S1**. The calculations of $X_1$ and $Y_1$. (a) Schematic of a Si disk excited by a plane wave. The height and diameter of the disk are $h = 630\ nm$ and $D = 1260\ nm$, respectively. The magnetic field distribution ($|B/B_0|$) pattern on the *x-y* plane of a TE WGM (*m* = 6) is also shown. (b) Magnetic polarizability $\alpha_M$ spectra of the TE WGM (*m* = 6). (c) Magnetic field enhancement spectra at a field antinode of the WGM. (d) Schematic of the Si disk excited by a MD emitter. (e) Purcell factor of the MD emitter around the TE WGM of *m* = 6. (f) Normalized $X_1(Y_1)$ at different locations of the cavity. *L* is the distance between the calculated location and the disk center.

The values of $X_1$ ($X_2$) and $Y_1$ ($Y_2$) can be calculated through different ways. For $X_1$ (or $X_2$), it is a represents the ratio between the magnetic field and the magnetic moment of a cavity according to its definition, namely, $X_1 = (B - B_0)/m_{cav}$. The magnetic moment can be expressed as $m_{cav} = \mu_0^{-1} \alpha_M B_0$. Therefore, one can rewrite $X_1$ as $X_1 = (B - B_0)/\mu_0^{-1} \alpha_M B_0 \approx B/B_0 \mu_0^{-1} \alpha_M$. The $B/B_0$ can be easily obtained directly by FDTD simulations [Figs. S1(a) and S1(b)]. The polarizability $\alpha_M$ can be calculated by a multipole decomposition method which is widely used to obtain the electromagnetic multipole moments of a cavity [S1,S2]. In this method, the electric fields around a cavity are required for further calculations, and these fields can also be obtained by FDTD simulations. The magnetic field enhancement at a WGM field antinode is $B/B_0 = 10 + 395i$. Based on the results in Figs. S1(b) and S1(c), the $X_1$ at a WGM field antinode is $7.53 \times 10^{14} - 1.78 \times 10^{13} i$. For $Y_1$ (or $Y_2$), it is not intuitive to calculate it similarly to that of $X_1$ as the cavity is much larger than an emitter. This value can be obtained indirectly through the calculation of the radiation of an emitter in a cavity. The decay rate enhancement (Purcell factor) of a MD emitter is

$$F_p^{(m)} = \frac{6\pi c^3}{\omega_0^3} \mu_0^{-1} |Y_1| \left|\frac{B}{B_0}\right|$$ according to Eq. (A4) in the text. In this expression the $B/B_0$ is the magnetic field enhancement of a cavity which is already obtained as shown in Fig. S1(b). The Purcell factor can be calculated by FDTD simulations by setting a MD emitter as the excitation source. Thus, $Y_1$ can be obtained correspondingly. As shown in Fig. S1(d), a MD emitter is placed in the Si cavity, which can excite the TE $m = 6$ WGM in the cavity. The Purcell factor is 45300 at $\lambda \approx 1371\,nm$, and the magnetic field enhancement is already obtained before ($B/B_0 = 10 + 395i$, see Fig. S1(c)). One can calculate $|Y_1| = 7.36 \times 10^{14}$ which agrees well with the $X_1$. Note that the simulations in Figs. S1(a-c) and S1(d, e) are different. The excitation source is a plane wave in Figs. S1(a-c) and a MD emitter in Figs. S1(d) and S1(e), respectively. The $X_1$ and $Y_1$ at

other locations of the cavity are also considered and they also agree well with each other [Fig. S1(f)]."

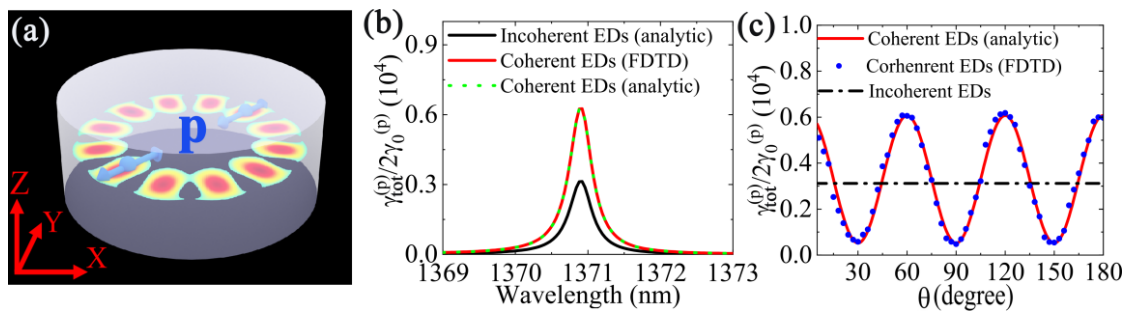

**FIG. S2.** (a) Schematic of the hybrid structure composed of a dielectric Si disk resonant cavity and two identical ED emitters. (b) The ratio of the total decay rate $\gamma_{tot}^{(p)}$ of the ED emitters in the Si disk to the decay rate $2\gamma_0^{(p)}$ of the same emitters in free space. (c) The ratio $\gamma_{tot}^{(p)}/2\gamma_0^{(p)}$ of the coherent and incoherent MD emitters as a function of $\theta$.

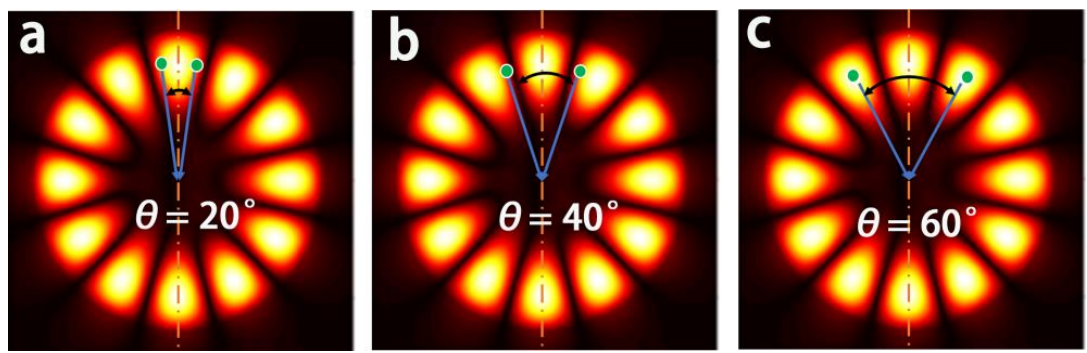

**FIG. S3.** (a, b, c) The patterns of magnetic fields of TE $m = 6$ WGM excited by MD emitters on the x-y plane. The distances between them are $\theta = 20°, \theta = 40°, \theta = 60°$, respectively.

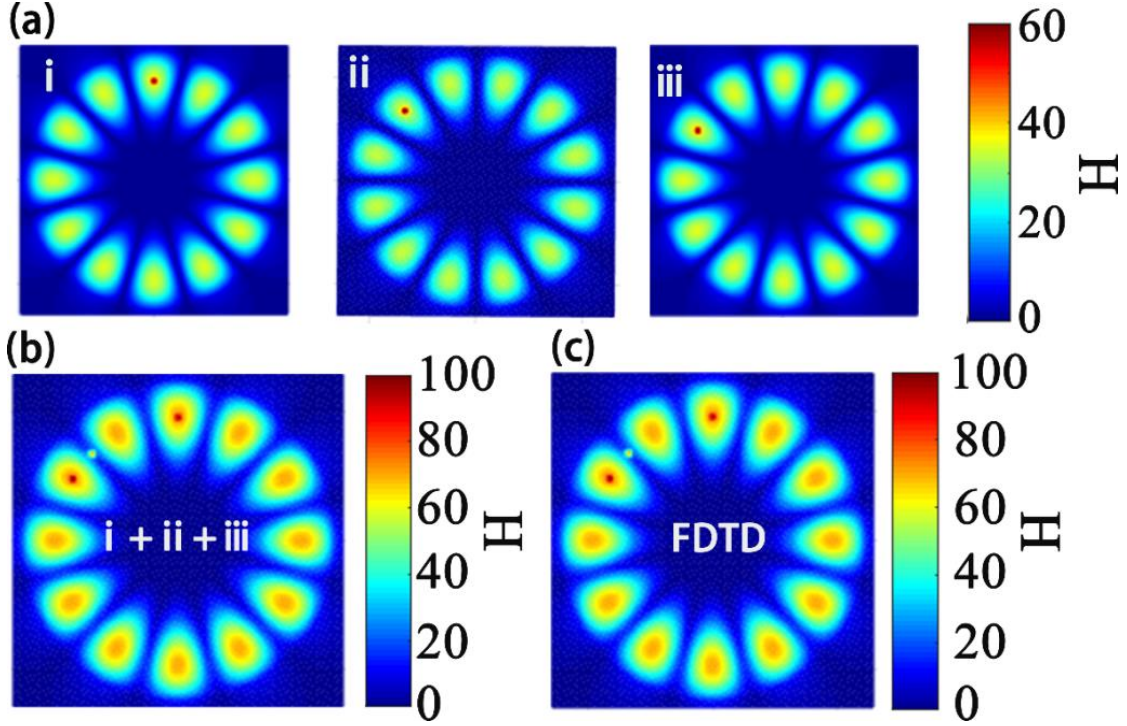

**FIG. S4.** (a) The patterns of magnetic field excited by each emitter located in different positions on *x-y* plane of the Si disk. (b) The magnetic field distribution carried out via superposition the value of the magnetic fields of i, ii, iii. (c) The pattern of magnetic field excited by three emitters obtained by numerical FDTD method, in which the locations of the three emitters correspond to the counterparts in i, ii, iii, respectively. The result is the same as that shown in Fig. S2(b), which means the magnetic field excited by multiple emitters can be taken as the superposition of the magnetic fields excited by each emitter individually.

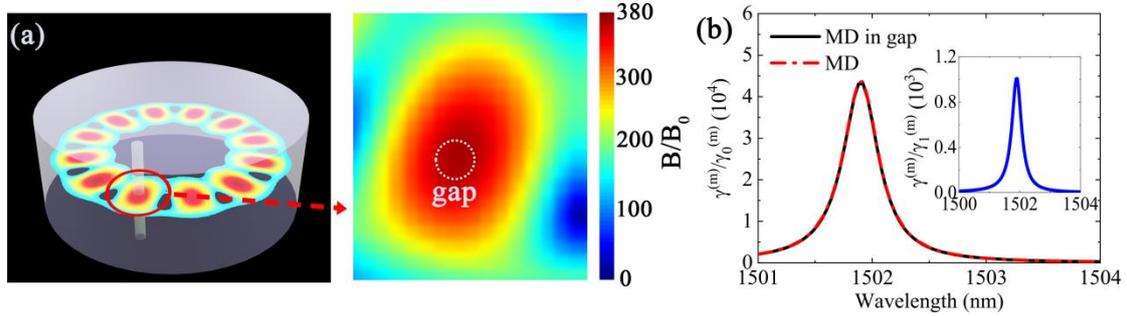

**FIG. S5.** （a）The magnetic field distributions of the TM $m = 6$ WGM on the *x-y* plane in the Si disk excited by the plane wave. A hole with a radius of 20 nm is introduced in the disk. (b) The ratios of $\gamma^{(m)}/\gamma_0^{(m)}$ in a solid disk and a disk with a small hole. The inset shows the ratio $\gamma^{(m)}/\gamma_1^{(m)}$ of the MD in a solid disk that calculated directly by the FDTD.